# Higher-Dimensional Bulk Wormholes and their Manifestations in Brane Worlds

Enrico Rodrigo

*Department of General Studies, Charles Drew University, Los Angeles, CA 90059*

Abstract

*There is nothing to prevent a higher-dimensional anti-de Sitter bulk spacetime from containing various other branes in addition to hosting our universe, presumed to be a positive-tension 3-brane. In particular, it could contain closed, microscopic branes that form the boundary surfaces of void bubbles and thus violate the null energy condition in the bulk. The possible existence of such micro branes can be investigated by considering the properties of the ground state of a pseudo-Wheeler-DeWitt equation describing brane quantum dynamics in minisuperspace. If they exist, a concentration of these micro branes could act as a fluid of exotic matter able to support macroscopic wormholes connecting otherwise distant regions of the bulk. Were the brane constituting our universe to expand into a region of the bulk containing such higher-dimensional macroscopic wormholes, they would likely manifest themselves in our brane as wormholes of normal dimensionality, whose spontaneous appearance and general dynamics would seem inexplicably peculiar. This encounter could also result in the formation of baby universes of a particular type.*

## I. INTRODUCTION

Brane world models (see [1] for a review) were originally developed in the late 1990s as a means of addressing the hierarchy problem – the seemingly unnatural discrepancy of sixteen orders of magnitude between the electroweak scale ($10^3$ GeV) and the Planck scale ($10^{19}$ GeV). They did so by modifying the traditional means of recovering the (3+1)-dimensional physics of our experience from the higher-dimensional spacetime natural to string theory. Rather than supposing that spacetime's extra dimensions are compactified to sizes on the order of a Planck length, they assumed instead that the compactification radii of these dimensions can be many orders of magnitude larger [2-4]. To prevent the disagreement with experiment that this assumption would have caused, these models confined all fields, with the exception of gravity, to a single macroscopic (3+1)-dimensional submanifold, or *3-brane*, of the higher-dimensional spacetime. This sort of confinement, a consequence of string theoretic effects, does not apply to gravity, because it is an inherent property of spacetime. While this confinement permits the compact dimensions of the higher-dimensional spacetime to be greatly enlarged, there is a limit. Currently, Cavendish experiments – those measuring the spatial dependence of the gravitational force – place this limit at a fraction of 1 millimeter [5]. The relative weakness of gravity (the huge size of the Planck mass) relative to electroweak interactions was thus explained as a consequence of gravity's dilution within these large extra dimensions to which non-gravitational fields have no access.

Randall and Sundrum [6,7] soon devised a novel variation on this theme in which they proposed what can be described in Wheeleresque fashion as a sort of "compactification without compactification". They allowed the bulk -- the higher-dimensional spacetime containing the 3-brane of the non-gravitational fields -- to possess non-compact extra dimensions. These dimensions were, however, compactified *in effect* by the negative curvature of the bulk, they having assumed the bulk to be an anti-de Sitter space. This curvature effected a sort of compactification by warping the higher dimensions so that spatial infinity was only a finite distance away from the 3-brane. This limited the distance over which gravity could effectively propagate out from the brane into the bulk. They showed this distance to diminish with increasing strength of the bulk's curvature.

If our world is actually a four-dimensional submanifold of the five-dimensional bulk most often considered, then the relevant Einstein equations are fundamentally 5D. Finding 5D solutions for which a 4D slice corresponds to the known solutions of 4D general relativity has proven to be difficult. The initial attempt, which consisted of creating a "black string" by stretching the Schwarzschild solution into the bulk, failed [8]. Its curvature turned out to be unbounded at its Cauchy horizon [8]. It is, moreover, generally unstable against perturbations [9]. Subsequent attempts focused on solving the Einstein equations induced on our 3-brane. These induced equations generalized the usual gravitational field equations by augmenting the right hand side with terms quadratic in the stress energy tensor and a term dependent on the Weyl tensor [10]. Solutions to these equations were usually obtained through three approaches. The first uses algebraic symmetries of the Weyl term to create an ansatz for it that corresponds to a fluid in the bulk [11-13]. Assuming the induced cosmological constant and the stress energy tensor to vanish on the brane, this approach yields solutions that depend upon the on-brane values of the functions that characterize the Weyl term ansatz. The second approach, which also assumes a vanishing on-brane cosmological constant and stress energy tensor, exploits the tracelessness of the Weyl term to reduce the Einstein equations to $R = 0$, where $R$ is the curvature scalar [14, 15]. Solutions are thus found without having to model the Weyl contribution to the effective stress energy. This contribution's on-brane values, however, remain unspecified parameters of the solution. The third approach embeds the brane in an asymptotically anti-de Sitter bulk spacetime – one containing a black hole or one in which the brane functions as a domain wall between regions of anti-de Sitter space [16-19]. This exact specification of the bulk spacetime induces precise values for the Weyl stress energy on the brane. However, the special symmetry of the exact bulk solution normally limits this approach to global descriptions of the brane. These are of use in investigating brane-world cosmologies.

Here, however, we are interested in effects that are noncosmological and local. Hence, we will suitably modify the third approach by eliminating its requirement that the brane lie on a submanifold of special symmetry with respect to the bulk solution. We choose the third approach because, unlike the other two, it is bulk-centric. It thereby attempts to solve the fundamental problem: obtaining *5D* solutions for which the 4D on-brane solutions are boundary values. The other two approaches are nevertheless of some relevance to the subject at hand. They both yield on-brane solutions that not only describe new sorts of black holes and naked singularities, but also contain traversable wormholes.

The discussion of wormholes in the context of brane worlds seems to have begun with the realization by Anchordoqui and Perez Bergliaffa that the Randall-Sundrum Version I construction – an anti-de Sitter bulk bounded by two branes, one of positive tension, the other negative – meets the formal definition of wormhole [20]. The negative-tension brane, which in RS Version I is identified with our universe, is a surface of minimal area – the definition of a wormhole throat. As noted by Anchordoqui and Perez Bergliaffa, RS Version I actually defines only half of a wormhole. It does not include a description of the presumably distant region of the bulk (perhaps another RS Version I construction) that would be encountered by a traveler in the bulk who were to somehow pass *through* the negative-tension 3-brane and thus traverse the wormhole's throat.

Unlike the considerations of Anchordoqui and Perez Bergliaffa, subsequent investigations of wormholes in brane worlds ignored higher-dimensional intra-bulk wormholes to concentrate on their 4D intra-brane counterparts. These included a description of (3+1)-dimensional wormholes throats as (2+1)-dimensional negative tension-branes [21], a numerical wormhole solution corresponding to an axially symmetric distribution (normal to our 3-brane) of matter in the bulk [22], and the aforementioned elucidation of the wormhole solutions to the contracted field equation $R=0$. The conviction that all solutions on the brane – including those describing wormholes -- are ultimately slices of a solution in the bulk drives us to return to higher-dimensional intra-bulk wormholes. The purpose of this note is to understand how wormholes on the brane can be induced by the presence of wormholes in the bulk. These on-brane wormholes could then be understood properly -- as four-dimensional manifestations of solutions to Einstein equations in the higher-dimensional bulk. An immediate obstacle to this effort is the apparent absence in the bulk of the exotic matter without which traversable wormholes cannot exist. I address this problem in the next section before attempting in a subsequent section to describe qualitatively the properties of an induced wormhole.

## II. BRANE-BASED EXOTIC MATTER IN THE BULK

### A. Procedure Delineated

Henceforth, we shall only consider RS Version II – which describes our universe as a positive-tension brane in an anti-de Sitter bulk in the *absence* of a "hidden" brane of opposite tension. To adhere faithfully to the Randall-Sundrum ethos is to eschew the possibility of matter within the bulk. I shall therefore assume the bulk to be solely inhabited by branes of various tensions, dimensionalities, and sizes. The bulk, besides hosting our own positive tension 3-brane, can in particular contain closed 3-branes that are miniscule compared to the macroscopic wormhole throats of interest. These small branes can, moreover, violate the null energy condition (NEC) in the bulk, as, for example, does the negative-tension brane of the original RS construction. An aggregation, then, of such NEC-violating micro 3-branes can serve as fluid of exotic matter able to support higher-dimensional intra-bulk wormholes. Although a 5D RS bulk may also be presumed to contain closed NEC-violating 2-branes and 1-branes (exotic gravitons), I shall only consider 3-branes and suppose them to be the leading contributors to exotic matter in the bulk.

I shall not consider the obvious possibility of macroscopic NEC-violating 3-branes that serve as the throats for macroscopic, thin shelled, intra-bulk wormholes. Doing so would require a theory of brane-brane interaction to describe the expansion of a dynamic 3-brane such as our universe into a region containing such an intra-bulk wormhole. This theory would in particular be required to describe the 3-brane's attempt to expand through the wormhole, where the brane would necessarily collide with the 3-brane at the wormhole's throat. By contrast, a 3-brane expanding into region of the bulk containing a wormhole supported by an exotic fluid of micro 3-branes would merely displace the fluid (which is mostly empty space) and thus avoid direct interaction with it. This displacement is of negligible consequence, because the *5D* volume of a 3-brane is precisely zero. I shall reserve for a future investigation the complicated issue of brane-brane interactions and their relevance to wormholes.

An exotic fluid composed of countless NEC-violating micro 3-branes is only possible if there exists a stable or metastable state in which such branes are at least a couple of orders of magnitude smaller than the macroscopic wormhole throats of interest. To determine whether this is the case, we will employ the thin-shell formalism [23, 24] to consider the quantum dynamics of a closed micro 3-brane that violates the bulk NEC. Rather than supposing that this micro brane is itself a wormhole throat, we shall instead regard it as a bubble surrounding the void. Like the negative-tension brane in the RS Version I construction, it is a boundary for the bulk spacetime. In other words the region surrounded by the micro 3-brane is *not* an interior piece of the anti-de Sitter bulk. Rather, this region is the void – it does not exist.

The expected reign of chaos and pandemonium at these bulk-void boundaries due, to the loss of predictability there, need not occur. As emphasized by Barceló and Visser [21], these boundaries are *Dirichlet* branes. As such the values of any fields capable of propagating within the bulk are perfectly determined at these timelike hypersurfaces. This ensures predictability within the bulk's natural Cauchy horizons.

The aforementioned thin-shell formalism, due presumably to the absence of energy or momentum fluxes across the thin shell, applies equally well to bulk-void as to bulk-bulk interfaces, as argued by Barceló and Visser. We need, then, to apply it to ascertain whether stable states corresponding to microscopic diameters exist for a negative-tension 3-brane enclosing a void bubble. Fortunately, Barceló and Visser have in practice already done so. Although they considered 2-branes embedded in a (3+1)-dimensional bulk (ordinary 4D spacetime), their results can be extended to the current case with little modification.

We will proceed according to the following recipe: 1) We will write down the metric for a thin-shell void bubble in the Reissner-Nordstrøm-anti-de-Sitter solution. 2) We will assume the bubble's radius to be dynamic and obtain the corresponding extrinsic curvature. 3) We will apply the Darmois-Israel junction conditions (the thin-shell formalism) [23, 25] and thus obtain a Friedmann-like "initial-value" equation. 4) We will supplement this equation with the First Law of Thermodynamics and an ansatz for the equation of the state, combine these equations, and thus obtain a single equation describing the dynamics of the bubble radius. 5) We will note that this equation may be derived from a Lagrangian with a kinetic term and a conservative potential term, and note further the existence of (meta) stable bubbles at the minima of the potential term. 6) We will perform a change of variable so that the kinetic term becomes purely quadratic in the proper time derivative of the new variable. 7) We will apply standard canonical quantization to the Lagrangian expressed in the new variable to obtain a Schrödinger-like equation. 8) We will use this equation to estimate the ground-state expectation value of the radius of quantum void bubbles. 9) We will infer the possible existence of such bubbles, if this expectation value exceeds the outer horizon of the black hole geometry from which the bubble is constructed.

## B. Classical Void Bubble

We begin with the n-dimensional Reissner-Nordstrøm metric in an asymptotically anti-de-Sitter background

$$ds^2 = -F(r)dt^2 + \frac{dr^2}{F(r)} + r^2 d\Omega_{n-2}, \tag{1}$$

where

$$F(r) = 1 - \frac{M_n}{r^{n-3}} + \frac{Q_n^2}{r^{2n-6}} - \Lambda_n r^2 \tag{2}$$

and

$$M_n \equiv \frac{16\pi G M}{(n-2)c^2 \Omega_{n-2}}, \quad Q_n^2 \equiv \frac{k_n Q^2 G}{2(n-2)(n-3)}\left(\frac{8\pi}{\Omega_{n-2}c^2}\right)^2, \quad \Lambda_n \equiv \frac{2\Lambda}{(n-1)(n-2)}. \tag{3}$$

Here $\Omega_{n-2}$ is the "surface area" of an (n-2)-dimensional unit hypersphere, c is the speed of light, $G_n$ is the n-dimensional gravitational constant, $k_n$ is the constant in the n-dimensional Coulomb potential, i.e.

$$\phi_g \approx -\frac{G_n M}{r^{n-3}} \qquad \phi_e \approx -\frac{k_n Q}{r^{n-3}}. \tag{4}$$

*M, Q, Λ* are the asymptotically observed mass, charge, and cosmological constant. The void bubble may be regarded as half of a wormhole. To construct it consider two disconnected manifolds, each of which contains a Reissner-Nordstrøm-anti-de-Sitter metric. Remove identical spheres, given by *r=a,* from both manifolds. Identify the resultant spherical boundary surfaces, and thus create a thin-shell wormhole. The dynamics of this wormhole are the same as those of a void bubble of equal radius [21]. Our analysis of the void bubble, then, will be a standard thin-shell wormhole analysis [26-32] in which we will discard the half of the wormhole exterior to our universe when we are done.

One proceeds by writing down a 4-vector for the a point on the dynamic surface *a(τ),* where *τ* is the proper time of an observer riding the surface. The vector $n^\mu$ normal to this surface can be used with the above metric to obtain the extrinsic curvature *K* of the thin shell,

$$K_{\mu\nu} = \frac{1}{2} n^\alpha \partial_\alpha g_{\mu\nu} . \tag{5}$$

Applying the Darmois-Israel junction conditions to this extrinsic curvature yields the equations of motion

$$\rho = -\frac{(n-2)c^4}{4\pi G_n a} \sqrt{F(a) + \frac{\dot{a}^2}{c^2}} , \tag{6}$$

$$p = \frac{c^4}{4\pi G_n a} \left[ (n-3)\sqrt{F(a) + \frac{\dot{a}^2}{c^2}} + a \frac{d}{da} \sqrt{F(a) + \frac{\dot{a}^2}{c^2}} \right] \tag{7}$$

where $\dot{a}$ is the derivative of the bubble radius with respect to proper time, *ρ* is the matter density within the thin shell, and *p* is the corresponding pressure. The junction conditions leave the signs leading the right sides of these equations undetermined. The choice shown, however, ensures that the NEC will be violated in the bulk irrespective of whether it is satisfied on the brane. The initial-value equation (6) together with the First Law of Thermodynamics applied to the shell,

$$\frac{d}{d\tau}\left(\rho a^{n-2}\right) = -p \frac{d}{d\tau} a^{n-2} , \tag{8}$$

renders eq. (7) superfluous. Despite its inadequacy we will for the sake of simplicity assume the usual equation of state for dilute matter, *p = wρ*. The First Law then becomes

$$\dot{\rho} + (n-2)(1+w)\frac{\dot{a}}{a}\rho = 0 , \tag{9}$$

which is immediately solved to obtain

$$\rho(\tau) = \rho(\tau_0)\left[\frac{a(\tau)}{a(\tau_0)}\right]^{-(n-2)(1+w)}. \tag{10}$$

Inserting this in eq. (6), squaring, and simplifying yields,

$$\rho_0^2 a_0^{2(n-2)(1+w)} = c^8 \left(\frac{n-2}{4\pi G_n}\right)^2 \left\{\frac{F(a)}{a^{2-2(n-2)(1+w)}} + \left[c^{-1}\frac{d}{d\tau}\left(\frac{a^{(n-2)(1+w)}}{(n-2)(1+w)}\right)\right]^2\right\} \tag{11a}$$

for $w \neq -1$, and

$$\rho_0^2 = c^8\left(\frac{n-2}{4\pi G_n}\right)^2\left[\frac{F(a)}{a^2} + \left(\frac{d\ln(a)}{cd\tau}\right)^2\right] \tag{11b}$$

for $w = -1$,

where $\rho_0 \equiv \rho(\tau_0)$ and $a_0 \equiv a(\tau_0)$. From this it is clear that the bubble's classical dynamics are governed by an effective potential function $V$ given by,

$$V(a) = \frac{F(a)}{a^{2-2(n-2)(1+w)}}, \tag{12}$$

and that the stable static solutions occur at this function's minima. Barceló and Visser [21] analyzed the case of a "clean" brane in four dimensions -- $w=-1$ and $n=4$. They found that, except for certain special values of the parameters $M$ and $Q$, stable solutions to (11b) exist. Here we will consider instead the case described by eq. (11a) with $n=5$ – void bubbles in a 5-dimensional bulk containing 3-branes whose internal equation of state corresponds in general to $w \neq -1$. Our focus on these "unclean" branes in 5D will be directed to their quantum dynamics.

We are in particular interested in quantum void bubbles as the fundamental constituents of wormhole-supporting exotic matter in the bulk. To find such bubbles, it helps to perform a change of variable from the bubble radius $a$ to a variable $q$ defined by

$$a = q^{\frac{1}{(n-2)(1+w)}}. \tag{13}$$

Inserting (13) in (11a) and multiplying by a constant factor gives

$$Mc^2\left(\frac{G_n M}{c^2}\right)^{\frac{2-2(n-2)(1+w)}{n-3}}\left[\frac{4\pi G_n}{(n-2)c^4}\right]^2 \rho_0^2 q_0^2 = \frac{Mc^2\left(\frac{G_n M}{c^2}\right)^{\frac{2-2(n-2)(1+w)}{n-3}}}{(n-2)^2(1+w)^2 c^2}\dot{q}^2 + U(q) \quad (14)$$

where

$$U(q) = Mc^2\left(\frac{G_n M}{c^2}\right)^{\frac{2-2(n-2)(1+w)}{n-3}} V(a(q)) \quad (15)$$

with

$$V(a(q)) = q^{2-\frac{2}{(n-2)(1+w)}} - M_n q^{2-\frac{n-1}{(n-2)(1+w)}} + Q_n^2 q^{2-\frac{2}{1+w}} - \Lambda_n q^2. \quad (16)$$

and $\quad q_0 \equiv q(a_0)$.

Interpreting eq. (14) as the "total energy" of an ordinary dynamical system -- one whose Lagrangian $L$ features a simple kinetic term and a conservative potential -- we may define the "momentum" p conjugate to q in the usual fashion,

$$L = \frac{M\left(\frac{G_n M}{c^2}\right)^{\frac{2-2(n-2)(1+w)}{n-3}}}{(n-2)^2(1+w)^2}\dot{q}^2 - U(q) \quad (17a)$$

$$p = \frac{\partial L}{\partial \dot{q}} = \frac{2M\left(\frac{G_n M}{c^2}\right)^{\frac{2-2(n-2)(1+w)}{n-3}}}{(n-2)^2(1+w)^2}\dot{q}, \quad (17b)$$

with which (14) becomes

$$H = \frac{(n-2)^2(1+w)^2}{4M\left(\frac{G_n M}{c^2}\right)^{\frac{2-2(n-2)(1+w)}{n-3}}} p^2 + U(q) \quad (18)$$

after re-labeling the product of integration constants on the left side of eq. (14) as

$$H \equiv M\left(\frac{G_n M}{c^2}\right)^{\frac{2-2(n-2)(1+w)}{n-3}}\left[\frac{4\pi G_n}{(n-2)c^3}\right]^2 \rho_0^2 q_0^2. \quad (19)$$

Note that $H$ has units of energy as a consequence of $\rho$ having units of energy per length$^{n-2}$ and $G_n$ having units of length$^{n-3} c^2$/mass.

## C. Quantum Void Bubble

Imposing the traditional commutation relation $[p,q] = -i\hbar$ and its realization, $p \to -i\hbar\, \partial/\partial q$, we have the operator

$$\hat{H} = -\frac{(n-2)^2(1+w)^2 \hbar^2}{4M\left(\dfrac{G_n M}{c^2}\right)^{\frac{2-2(n-2)(1+w)}{n-3}}} \frac{\partial^2}{\partial q^2} + U(q), \tag{20}$$

whose eigenvectors $\psi(q)$ and eigenvalues $E$ satisfy

$$\hat{H}\psi = -b(n,w)\hbar^2 \frac{\partial^2 \psi}{\partial q^2} + U(q)\psi = E\psi \tag{21a}$$

with

$$b(n,w) \equiv \frac{(n-2)^2(1+w)^2}{4M\left(\dfrac{G_n M}{c^2}\right)^{\frac{2-2(n-2)(1+w)}{n-3}}} \tag{21b}$$

This may be regarded as a sort of poor man's minisuperspace Wheeler-DeWitt equation for the (n-2)-spherical void bubble. Of course, unlike eq. (21a), the actual Wheeler-DeWitt equation stems from the Hamiltonian constraint and not from the square of the initial-value equation (6). Moreover, the true Wheeler-DeWitt equation features in this case nonlocal differential operators [28, 33]. Indeed, such operators would have arisen in (21a) were it expressed in terms of coordinate time instead of the proper time of an observer on the void bubble, or were it constructed from the initial-value equation itself instead of its square. Nevertheless, eq. (11a) – the square of the initial-value equation with an equation of state and the First Law inserted – exhausts the dynamical content of the bubble model. Hence, its corresponding Lagrangian is as valid a candidate as any for the usual canonical quantization despite the implicitly chosen gauge (lapse and shift) inherent in the above procedure. [See [34] and [35] for a similar approach applied to another initial-value equation, the Friedmann Equation, and [36] for an application to wormholes.]

Normally, we would employ the WKB approximation to find solutions to eq. (21a). However, we are only interested in whether a stable ground state exists and in the approximate expectation value of the corresponding bubble radius. The cheapest way to determine this is to apply the uncertainty principle. Replacing $p$ with $q^{-1}\hbar/2$ in eq. (18) creates a function $H(q)$, given by

$$H(q) = b(n,w)\frac{\hbar^2}{4q^2} + U(q),$$

whose minimum, if present, indicates the existence of a stable quantum ground state. For the sake of convenience we scale $H(q)$ to define $\tilde{H}(q)$ as

$$\tilde{H}(q) \equiv \frac{4b(n,w)}{(n-2)^2(1+w)^2 c^2} H(q)$$

$$= \frac{b^2(n,w)\hbar^2}{(n-2)^2(1+w)^2 c^2} q^{-2} + q^{2-\frac{2}{(n-2)(1+w)}} - M_n q^{2-\frac{n-1}{(n-2)(1+w)}} + Q_n^2 q^{2-\frac{2}{1+w}} - \Lambda_n q^2. \quad (22)$$

Because we are not interested in enforcing the weak energy condition at the bubble's throat, we need not be concerned that our warp-compactified extra dimension might restrict the bubble's radius [37] taken to be a minimum of $\tilde{H}$. This concern is particularly unwarranted given that the 5D case henceforth considered happens to be exempt from such restrictions [37]. Any limitations, then, on the radii of the bubbles that we will consider will be due solely to the requirement that these radii exceed the event horizons of the black hole geometries from which the bubbles are constructed.

## D. Application to 5-D Bulk

Specializing to the case of a 5-dimensional bulk, we have

$$\tilde{H}(q) = A_w q^{-2} + q^{\frac{4+6w}{3+3w}} - M_5 q^{\frac{2+6w}{3+3w}} + Q_5^2 q^{\frac{2w}{1+w}} - \Lambda_5 q^2, \quad (23)$$

where

$$A_w \equiv \frac{b^2(5,w)\hbar^2}{9(1+w)^2 c^2} = \left[\frac{3(1+w)\hbar}{4Mc}\left(\frac{G_5 M}{c^2}\right)^{2+3w}\right]^2. \quad (24)$$

Consider this function $\tilde{H}(q|w)$ for a few values of $w$:

$$\tilde{H}(q|2/3) = A_{2/3} q^{-2} + q^{8/5} - M_5 q^{6/5} + Q_5^2 q^{4/5} - \Lambda_5 q^2 \quad (25a)$$

$$\tilde{H}(q|1/3) = A_{1/3} q^{-2} + q^{3/2} - M_5 q + Q_5^2 q^{1/2} - \Lambda_5 q^2 \quad (25b)$$

$$\tilde{H}(q|0) = A_0 q^{-2} + q^{4/3} - M_5 q^{2/3} + Q_5^2 - \Lambda_5 q^2 \quad (25c)$$

$$\tilde{H}(q|-1/3) = A_{-1/3} q^{-2} + q - M_5 + Q_5^2 q^{-1} - \Lambda_5 q^2 \quad (25d)$$

$$\tilde{H}(q|-2/3) = A_{-2/3} q^{-2} + 1 - M_5 q^{-2} + Q_5^2 q^{-4} - \Lambda_5 q^2 \quad (25e)$$

where

$$M_5 = \frac{8G_5 M}{3\pi c^2}, \qquad Q_5^2 = \frac{4k_5 G_5 Q^2}{3\pi^2 c^4}, \qquad \Lambda_5 = \frac{\Lambda}{6}. \tag{26}$$

It is clear from inspection that minima are likely to exist for each value $w$. What is unclear is whether these minima exist outside of the exterior event horizon of the Reissner-Nordstrøm-anti-de-Sitter solution, whose value is a root of

$$F(r_h) = 1 - \frac{M_5}{r_h^2} + \frac{Q_5^2}{r_h^4} - \Lambda_5 r_h^2 = 0. \tag{27}$$

This is essentially a cubic equation in $r_h^2$. Its three real roots correspond to the cosmological de Sitter horizon (which goes to infinity as a *positive* $\Lambda$ goes to zero), the inner Cauchy horizon of the Reissner-Nordstrøm-anti-de-Sitter black hole, and its external event horizon. We seek minima of the functions (25) exterior to the last of these. We shall determine numerically the smallest mass $M$ at which the locations of the minima of the function $\tilde{H}$ are exterior to the outer event horizon. We shall thus demonstrate the existence of a stable bubble ground state and approximate its mass and size for given values of its charge and the cosmological constant.

Recent experiments show Newton's $1/r^2$ law of gravitational attraction to hold for mass separations $s$ as small as $s = 0.1$ mm [5]. This limits the energy scale $M_{P5}$ of the RS Version II (single brane) bulk, according to [1]

$$M_{P5} > \left( \frac{M_{P4}^2 \hbar}{sc} \right)^{\frac{1}{3}} \sim 10^5 \, Tev/c^2, \tag{28}$$

and the bulk cosmological constant by

$$\Lambda < -\left[ \frac{(n-2)(n-3)}{s^2} \right]_{n=5} = -\frac{6}{s^2} = -6 \times 10^4 \, cm^{-2}, \tag{29}$$

where $M_{P4}$ is the on-brane Planck mass. The bulk gravitational constant $G_5$ corresponding to eq. (28) is

$$G_5 = \left[ \frac{\hbar^{n-3} c^{5-n}}{M_{Pn}^{n-2}} \right]_{n=5} = \frac{\hbar^2}{M_{P5}^3} = 2 \times 10^{-7} \, cm^4 \, \sec^{-2} \, gm^{-1}, \tag{30}$$

where $M_{P5} = 10^5$ TeV/$c^2$ Because we are after a quantum wormhole, I will suppose that its charge, whenever it is nonzero, consists of a single quanta of usual value

$$|Q| = 4.80 \times 10^{-10} \, statcoulomb, \tag{31}$$

i.e. the charge of the electron. I will suppose further that coupling constant $k_n$ of eq. (4) retains in the bulk its on-brane value of 1 in CGS units.

We shall consider each of the cases shown in eqs.(25) in turn.

## E. Specific Cases of Brane Matter

*1. w = -2/3, Q = 0: Uncharged Quintessence*

Because we have assumed $\Lambda_5$ to be negative, eq. (25e) makes it clear that $\tilde{H}(q)$ possesses minima at q > 0 only if $A_{-2/3} > M_5$. This puts an upper limit on the Arnowit-Deser-Misner (ADM) mass [56] of the void bubble. This mass is somewhat more tightly constrained by the requirement that the estimated expectation value $a_0$ of the bubble's ground-state radius be exterior to its event horizon $r_h$,

$$a > r_h \tag{32}$$

where

$$a = \left(\frac{A_{-2/3} - M_5}{-\Lambda_5}\right)^{1/4} = \left\{\left[\frac{8G_5 M}{3\pi c^2} - \left(\frac{\hbar}{4Mc}\right)^2\right]\Lambda_5^{-1}\right\}^{1/4} \tag{33}$$

$$r_h = \left(\frac{1 - \sqrt{1 - 4M_5\Lambda_5}}{2\Lambda_5}\right)^{1/2} \doteq \sqrt{M_5} \quad \text{for } M_5\Lambda_5 \ll 1. \tag{34}$$

Requiring *a* to be real, we obtain the upper limit

$$M < \left[\frac{3\pi}{8G_5}\left(\frac{\hbar}{4}\right)^2\right]^{1/3} \equiv M_{max} \doteq 7.45 \times 10^{-17} \, gm, \tag{35}$$

and the corresponding lower limit

$$a > \left(\frac{8G_5 M_{max}}{3\pi c^2}\right)^{1/2} = \left(\frac{2G_5\hbar}{3\pi c^3}\right)^{1/3} = 1.18 \times 10^{-22} \, cm. \tag{36}$$

In other words, ground-state quantum void bubbles in this case can be no more massive than an aggregate of about $10^7$ nucleons and must be at least $10^{11}$ Planck lengths wide. Moreover, the less massive the bubble, the larger it is. Without speculating on possible mechanisms for bubble formation, one might suppose such processes to be highly localized. These would favor, then, the formation of relatively massive bubbles, which are necessarily small and localized, over lighter bubbles extended over a large area.

While eq.(33) permits arbitrarily large bubbles of vanishing mass, we might suppose, therefore, that these are unlikely to form and are poorly represented in the bulk.

We should note further that the condition (32) on the ground-state *expectation value* of the void bubble's radius (where the ground state is that corresponding to the lowest eigenvalues $E$ of eq.(21a)) does not guarantee a stable state. Quantum fluctuation would collapse such state to a black hole at a rate that would depend inversely on the separation between the black hole horizon and the expectation value of the ground-state radius. We will always find, however, that this separation increases with decreasing mass $M$. Hence, metastable states that endure for arbitrarily long periods – states that are effectively stable -- may be supposed to exist.

## 2. $w = -2/3$, $Q \neq 0$: Charged Quintessence

Again our objective is to find a mass at which the expectation value of the ground-state bubble radius exceeds the external horizon of the Reissner-Nordstrom-anti-de-Sitter black hole geometry from which the bubble is constructed. Choosing $Q$ to be a small integral multiple of the charge of the electron, we find that the polynomial equation that determines the bubble radius and the cubic equation that determines the locations of the horizons can be approximated by ignoring their terms of the highest degree. The bubble radius $a$, the location at which $\tilde{H}(q|-2/3)$ is minimal, then solves

$$0 = (A_{-2/3} - M_5)q^2 + 2Q_5^2 + \Lambda_5 q^6 \doteq (A_{-2/3} - M_5)q^2 + 2Q_5^2 \doteq -M_5 q^2 + 2Q_5^2 \quad (37)$$

or

$$a = q \doteq \sqrt{\frac{2Q^2}{M_5}} \quad (38)$$

where the small value of $\hbar$ also renders $A_{-2/3}$ negligible in comparison to $M_5$ in the vicinity of the roots, and $a=q$ as a consequence of eq.(13). From the eq.(27), similarly approximated by discarding the term of highest degree, we may obtain the location of the black hole horizons as

$$0 = r_h^4 - M_5 r_h^2 + Q_5^2 - \Lambda r_h^6 \doteq r_h^4 - M_5 r_h^2 + Q_5^2 \quad (39)$$

or

$$r_h \doteq \left( \frac{M_5 \pm \sqrt{M_5^2 - 4Q_5^2}}{2} \right)^{1/2}. \quad (40)$$

That the roots $r_h$ must be real imposes a minimum value on $M_5$ and therefore $M$,

$$M_5 \geq 2Q_5 \quad (41a)$$

or

$$M \geq \sqrt{\frac{3kQ^2}{4G_5}} = (9.3 \times 10^{-7} gm)n_Q. \qquad (41b)$$

where $Q$ is $n_Q$ times the charge on the electron. For any $M$ above this minimum value, the larger of horizons given by (40) exceeds the bubble radius given by (38). The bubbles in this case do not exist. Interestingly, when $M$ is at its minimum, the bubble radius and the outer horizon coincide at the value

$$a = r_h = \sqrt{Q_5} = \left(\frac{4kG_5Q^2}{3\pi^2 c^4}\right)^{1/4}. \qquad (42)$$
$$= (9.4 \times 10^{-18} cm)\sqrt{n_Q}$$

Such a minimal bubble would, to observers near the horizon, appear to collapse immediately into a 5-dimensional black hole. However, "distant" observers could, presumably, note the existence of a stable quantum void bubble as a consequence of its collapse being frozen by gravitational time dilation. One could also argue that for sufficiently low mass/charge ratios, horizons do not exist. Void bubbles would not, then, be endangered by them. However, it is unclear whether the unphysical nature (in the vicinity of the origin) of the Reissner-Nordstrom geometry with low $M/Q$ would permit the construction of a disease-free void bubble. We shall assume, therefore, that M/Q is sufficiently large to permit the existence of horizons. Given that void bubble's radius does not clear the outer horizon, we conclude -- gravitational time dilation notwithstanding -- that a charged void bubble does not in this case exist.

### 3. w = -1/3, Q = 0: Uncharged Quasi-Quintessence

In this case we find that the expectation value of the bubble's ground-state radius does not depend on its ADM mass. Its fixed value is given by

$$a \doteq \left[\frac{1}{2}\left(\frac{\hbar G_5}{c^3}\right)^2\right]^{1/6} = 1.8 \times 10^{-22} cm. \qquad (43)$$

Requiring this to exceed the outer horizon, given by eq.(40), imposes an upper limit on the bubble's mass

$$M < \frac{3\pi \hbar^{2/3}}{2^{10/3} G_5^{1/3}} = 1.7 \times 10^{-16} gm. \qquad (44)$$

These bubbles can be up to one hundred million times more massive than a nucleon but only one billionth as large.

### 4. *w = -1/3, Q ≠ 0: Charged Quasi-Quintessence*

As in the case of *w=-2/3* for a nonzero charge, we find that eqs.(41) and (42) hold. Here, however, the bubble radius does not depend on the bubble's mass. As before, this radius lies within the exterior horizon of black hole geometries whose masses exceed the Reissner-Nordstrom minimum. Again we conclude that charged bubbles do not in this case exist.

### 5. *w = 0, Q = 0: Uncharged Dust*

Quantum Reissner-Nordstrom-Anti-de-Sitter void bubbles, whose 3-brane boundaries are characterized by uncharged dust, cannot exist. To see this, note that the location of the outer horizon is given by (34), but the bubble's radius turns out to be a constant fraction of this value,

$$a \doteq \sqrt{\frac{M_5}{2}} \doteq \frac{r_h}{\sqrt{2}}. \tag{45}$$

This violates the condition (32) required for the bubble's existence.

### 6. *w = 0, Q ≠ 0: Charged Dust*

As in the case of charged matter for *w=-2/3* and *w=-1/3*, the bubble's radius lies within the exterior horizon of its Reissner-Nordstrom-anti-de-Sitter geometry. Again this radius and the horizon only coincide when the bubble's mass has the minimum value allowed by its charge. Again we conclude that charged void bubbles, this time with *w=0*, are unlikely to exist. In this case, however, the bubble's radius does not decrease with increasing mass (as it does for w=-2/3), or remained fixed (as it does for w=-1/3). It increases with increasing mass, though not enough to allow it to exceed the horizon, which also increases at a similar rate.

### 7. *w = 1/3, Q = 0: Uncharged Relativistic Gas*

Finding the minima of (25b) and using eq. (13), the bubble radius $\underline{a}$ is the positive real root of

$$-2\left(\frac{\hbar G_5^{\,3} M^2}{c^7}\right) + \frac{3}{2}a^{14} - M_5 a^{12} - 2\Lambda_5 a^{16} = 0. \tag{46}$$

For numerically determined values of *M* near which the bubble radius equals the outer horizon of its Reissner-Nordstrom-anti-de-Sitter geometry, (46) may be approximated by

$$-2\left(\frac{\hbar G_5^{\,3} M^2}{c^7}\right) + \frac{3}{2}a^{14} - M_5 a^{12} = 0, \tag{47}$$

which suggests the iterative approximation scheme,

$$a_{n+1} = \left\{ \frac{2}{3} \left[ M_5 a_n^{12} + 2\left( \frac{\hbar G_5^3 M^2}{c^7} \right)^2 \right] \right\}^{1/14}, \tag{48}$$

in which the first term $a_0$ is obtained by setting the smallest term in eq.(47) – the constant term – to zero, yielding

$$a_0 = \sqrt{\frac{2}{3} M_5} \tag{49}$$

and

$$a \approx a_1 = \left\{ \frac{2}{3} \left[ \left(\frac{2}{3}\right)^6 \left(\frac{8 G_5 M_5}{3\pi c^2}\right)^7 + 2\left(\frac{\hbar G_5^3 M_5^2}{c^7}\right)^2 \right] \right\}^{1/14}. \tag{50}$$

Requiring this bubble radius to exceed the outer horizon approximated by (34), gives

$$M < \left[ \frac{2\left(\frac{3\pi}{8}\right)^7}{\frac{3}{2} - \left(\frac{2}{3}\right)^6} \hbar^2 G_5^{-1} \right]^{1/3} = 3 \times 10^{-16} \, gm. \tag{51}$$

Inserting this maximum bubble mass into (50), limits the bubble radius to

$$a < 2.6 \times 10^{-22} \, cm. \tag{52}$$

These void bubbles, whose 3-branes are characterized by uncharged relativistic gas, are similar in mass and size as those characterized by uncharged quintessence and quasi-quintessence.

## 8. $w = 1/3$, $Q \neq 0$: Charged Relativistic Gas

Again we find that bubble radius is never exterior to the outer horizon. As with the other charged bubbles, the radius coincides with the horizon when the bubble's mass assumes the minimal value given by eq.(41b). This is merely the mass at which the effectively Reissner-Nordstrom geometry becomes that of an extremal black hole. Void bubbles, then, whose boundary matter consists solely of charged relativistic gas do not appear to exist.

## 9. $w = 2/3$, $Q = 0$: Uncharged Quasi-Stiff Matter

Repeating the familiar procedure of estimating the minima of $\tilde{H}(q)$ by dropping terms negligible in their vicinity we see that requirement (32) leads to an ADM mass limit of

$$M \leq \left(\frac{3\pi}{8}\right)^3 \left(\frac{4}{3}\right)^{1/3} \left(\frac{5}{4}\hbar\right)^{2/3} G_5^{-1/3} \equiv M_{max} \approx 4 \times 10^{-16} \, gm. \quad (53)$$

The estimated expectation value $a_0$ of the void bubble's ground-state radius is correspondingly limited as

$$a \leq \left(\frac{25\pi\hbar^2}{32c^{16}} G_5^{7} M_{max}^{5}\right)^{1/16} \approx 2.7 \times 10^{-22} \, cm. \quad (54)$$

### 10. $w = 2/3$, $Q \neq 0$: Charged Quasi-Stiff Matter

As in the previous charged cases the $\Lambda$ term of $\tilde{H}(q)$ may be ignored in the vicinity of its minima. It again becomes clear that the bubble radius does not clear the horizon. Void bubbles whose surrounding branes are dominated by charged quasi-stiff matter cannot exist.

## F. Void Bubbles in Summary

It appears, then, that metastable quantum void bubbles can exist within a five-dimensional bulk. Their stability against runaway expansion is assured by the negativity of the bulk's cosmological constant. Their stability against collapse into black holes can be arbitrarily improved by lowering their masses (or, perhaps, by a more judicious application of the uncertainty principle, one that considers fluctuations in the radius about the classical minima of the effective potential rather than the coarse expedient of considering only fluctuations about zero). They are uncharged and can be much more massive than common elementary particles, while being much smaller. Specifically, they are no larger than $10^{11}$ Planck lengths, and no more massive than $10^{-11}$ Planck masses ($10^5$ TeV). Like conventional models of the primordial universe, they are, moreover, characterized by energy densities that diverge with decreasing volume. As is the case for stress energy at a wormhole's throat [38], these energy densities are independent of their apparent (i.e. ADM) mass.

## III. Implications and Speculations

### A. Macroscopic Bulk Wormholes

Having established the possibility of closed and quantum mechanically stable micro 3-branes (void bubbles) in the bulk, we now have the constituents for an exotic fluid able to support wormholes in the bulk. The notion of wormholes inhabiting the bulk might seem speculative or even fanciful. Consider, however, that much of the brane-world literature, motivated by the holographic principle and AdS/CFT correspondence, explores the ramifications of black holes in the bulk [see, for example, 17-19, 39-41]. The origins of these black holes -- which, due to the prohibition of off-brane matter, could not be the result of stellar collapse -- is seldom, if ever, discussed. It suffices to know that they are solutions to the gravitational field equations in the bulk and are therefore theoretical possibilities worthy of investigation. The possible existence of stable void bubbles permits us to consider bulk-dwelling wormholes in the same vein. Study of such wormholes is further justified by the recent realization that they are closely related to black holes, as it is possible in certain circumstances to convert wormholes into black holes and vice versa [42-44]. Moreover, the concept of a fluid of micro void bubbles is not dissimilar to that of a fluid of mini black holes in the bulk, which were considered within the last few years as a possible explanation for dark matter [45].

Because it appears that exotic matter can support *stable* wormholes in 4D spacetime [25, 46], we will assume – without taking the trouble to extend the four-dimensional analysis – that such wormholes are possible as well in spacetimes of higher dimension. This appears to be the case in five dimensions, at least in the presence of a Gauss-Bonnet term [47]. We shall, moreover, assume the possible existence of stable higher-dimensional wormholes even when the higher-dimensional spacetime is asymptotically anti-de Sitter. [See [47, 48] for investigations of wormholes with $\Lambda \neq 0$.]

I have hitherto been cavalier in my use of the adjectives "microscopic" and "macroscopic". A macroscopic wormhole should be understood to be one whose radius is on the order of the length scale determined by the bulk cosmological constant $\Lambda$, i.e. $|\Lambda|^{-1/2}$, or larger. A microscopic wormhole is one that is not macroscopic. Given that we have assumed this scale to be on the order of 0.1 mm, these adjectives nearly retain their literal meanings.

**B. Encounter between a 3-Brane and a Bulk Wormhole**

Recently, there has been a flurry of interest in the study of encounters between branes and black holes in the bulk [45, 49-52]. This follows the exploitation of the duality between certain gauge theories resembling high-temperature QCD and the gravitational interaction between branes and a higher-dimensional AdS black hole. In 2003 Shuryak pointed out that the unexpectedly low viscosities of high-temperature quark-gluon plasmas, inferred from data produced at the Relativistic Heavy Ion Collider, match those calculated using the duality, i.e. calculated by considering the encounter between a brane and black hole in a higher-dimensional bulk [53]. The most salient feature of this encounter, when it is close, is that the black hole in the bulk induces a black hole in the brane [49]. Here we consider instead an encounter between a brane and a wormhole in the bulk. The black hole case suggests that such an encounter will have an analogous result: A wormhole in the bulk will induce a wormhole in the brane.

Figure 1. Shows the intuitive basis for this speculation. As shown, one might imagine the brane to envelope the wormhole throat, rather as a latex sheet might envelope a bowling ball. If such envelopment is partial, it would result in the formation in the brane of a wormhole leading to a "pocket" universe -- one that remains connected to and consumes volume from its parent. If the envelopment is total -- i.e. if the wormhole induced in the brane pinches off -- it would result in a baby universe whose hypersurface would initially match that of the wormhole's throat. The details of the actual envelopment would depend on the Lagrangian chosen for the brane and, less critically, on the metric of the spherically symmetrical wormhole. We leave this calculation for a future investigation.

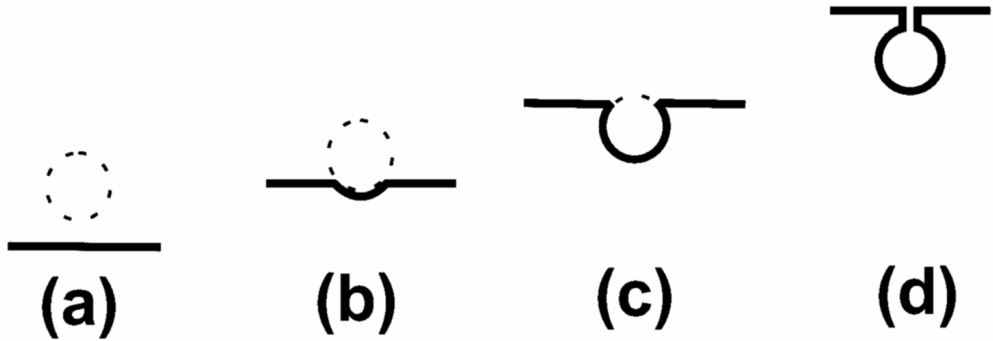

**Figure 1. Encounter between a brane and a bulk wormhole.** (a) A 3-brane (solid curve) encounters a preexisting wormhole in the bulk (dashed curve outlines its throat). (b,c) The brane envelopes the bulk wormhole's throat. (d) This envelopment induces a wormhole in the 3-brane that leads to a tiny pocket universe.

## IV. CONCLUSION

In the context of a brane-world scenario, I have considered the possibility of wormholes in the bulk and their likely manifestations in any large-scale brane that encounters them. In order for such wormholes to exist, the bulk must possess matter that violates the null energy condition. This need not imply the existence in the bulk of an oft-considered dilatonic scalar field [54, 55] that would violate the spirit (though not the letter) of the brane-world proscription of off-brane matter. This matter can instead consist of a fluid of closed microscopic 3-branes. This would only be possible if such 3-branes were quantum mechanically stable and possessed of exotic matter. Modeling such a 3-brane as a matter dominated domain wall in the Reissner-Nordstrom-anti-de-Sitter solution, I have derived its equation of motion using the standard thin-shell formalism and have canonically quantized a system whose extremized Lagrangian generates these equations. This procedure coupled with the uncertainty principle leads to an effective potential, whose minima correspond to quantum ground states. I have shown that such minima exist for several choices for the equation of state of on-brane matter and have focused specifically on branes whose matter content dominates their intrinsic tension (i.e. $p \neq -\rho$) The micro 3-branes that these minima describe are no larger than $\sim 10^{-22}$ cm. Their masses cannot exceed ~$10^5$ TeV (~$10^{-16}$ gm), lest they collapse and become black holes. It appears, moreover, that these branes may not be charged or be comprised solely of "dust" – matter without pressure.

In order to ensure the violation by these micro branes of the null energy condition in the bulk, I have modeled them as void bubbles – closed boundaries between the bulk and nothingness.

Having argued for the possibility of wormholes in the bulk supported by a fluid of exotic void bubbles, I have briefly considered an encounter between such a wormhole and our universe. An analogy with the encounter between bulk black holes and branes together with an intuitive understanding of the geometry of the encounter suggests that wormholes in the bulk will induce wormholes in the brane. A calculation of the details of the encounter merits a separate investigation.

Just as bubbles in champagne alter its character in interesting ways, so, it appears, do bubbles in the bulk.

# **REFERENCES**

[1] R. Maartens, *Living Rev. Rel.* **7**, 7 (2004), gr-qc/0312059
[2] N. Arkani-Hamed, S. Dimopoulos, and G. Dvali, *Phys. Lett. B* **429**, 263 (1998)
[3] N. Arkani-Hamed, S. Dimopoulos, and G. Dvali, *Phys. Rev. D* **59**, 086004 (1999)
[4] I. Antoniadis, N. Arkani-Hamed, S. Dimopoulos, and G. Dvali, *Phys. Lett. B* **436**, 257 (1998)
[5] J. Long, H. Chan, A. Churnside, E. Gulbis, M. Varney, and J. Price, *Nature*, **421**, 922 (2003)
[6] L. Randall and R. Sundrum, *Phys. Rev. Lett.* **83**, 3370 (1999)
[7] L. Randall and R. Sundrum, *Phys. Rev. Lett.* **83**, 4690 (1999)
[8] A. Chamblin, S. Hawking, and H. Reall, *Phys. Rev. D* **61**, 065007 (2000)
[9] R. Gregory, *Class. Quantum Grav*. **17**, L125-L131 (2000)
[10] T. Shiromizu, K. I. Maeda, and M. Sasaki, *Phys. Rev. D* **62**, 024012 (2000)
[11] N. Dadhich, R. Maartens, P. Papadopoulos, and V. Rezania, *Phys. Lett. B* **487**, 1 (2000)
[12] R. Maartens, *Phys. Rev. D* **62**, 084023 (2000)
[13] C. Germani and Roy Maartens, *Phys. Rev. D* **64**, 124010 (2001)
[14] R. Casadio, A. Fabbri, and L. Mazzacurati, *Phys. Rev. D* **65**, 084040 (2002)
[15] K.A. Bronnikov and S. Kim, *Phys. Rev. D* **67**, 064027 (2003)
[16] D. Birmingham, *Class. Quantum Grav*. **16**, 1197 (1999)
[17] P. Kraus, *JHEP* **9912**, 011 (1999)
[18] C. Barceló and M. Visser, *Phys.Lett. B* **482** (2000) 183-194
[19] J. Gregory and A. Padilla, *Class. Quantum Grav*. **19**, 4071 (2002)

[20] L. Anchordoqui and S. Perez Bergliaffa, *Phys. Rev. D* **62**, 067502 (2000)
[21] C. Barceló and M. Visser, *Nucl. Phys. B* **584**, 415 (2000)
[22] M. La Camera, *Phys. Lett. B* **573**, 27 (2003)
[23] G. Darmois in *Mémorial des Sciences Mathetmatiques* XXV, Gauthier-Villars, Paris (1927), Chapter V
[24] W. Israel, *Nuovo Cimento* **44B**, 1 (1966); *Nuovo Cimento* **48B**, 463 (1967)
[25] F. S. N. Lobo, *Phys.Rev. D* **71**, 124022 (2005)
[26] M. Visser, *Nucl. Phys. B* **328**, 203 (1989)
[27] M. Visser, *Phys. Rev. D* **43**, 402 (1991).
[28] M. Visser, *Lorentzian Wormholes*, AIP Press, Woodbury, NY (1996), Chapter 24
[29] D. Hochberg and T. W. Kephart, *Phys. Rev. Lett.* **70**, 2665 (1993).
[30] K. Narahara, Y. Furihata, and K. Sato, *Phys. Lett. B* **336**, 319 (1994).
[31] S. W. Kim, *Phys. Lett. A* **166**, 13 (1992)
[32] S. W Kim, H. Lee, S. K. Kim and J. Yang, *Phys. Lett. A* **183**, 359 (1993)
[33] M. Visser, *Phys. Lett. B* **242**, 24 (1990)
[34] M. Novello, J. M. Salim, M. C. Motta da Silva, R. Klippert, *Phys. Rev. D* **54**, 6202 (1996)
[35] E. Elbaz, M. Novello, J. Salim, M. Motta da Silva, R. Klippert, *Gen. Rel. Grav.* **29**, 481 (1997)
[36] I. H. Redmount and W. M. Suen, *Phys. Rev. D* **49**, 5199 (1994)
[37] A. DeBenedictis and A. Das, *Nucl. Phys. B* **653**, 279 (2003)
[38] M. Visser, (1996) *op. cit.*, Chapter 15
[39] E. Verlinde, hep-th/0008140 (2000)
[40] I. Savonije and E. Verlinde, *Phys. Lett. B* **507**, 305 (2001)
[41] R. G. Cai, *Phys. Rev. D* **63**, 124018 (2001)
[42] H. Koyama and S. A. Hayward, *Phys. Rev. D* **70**, 084001 (2004)
[43] S. A. Hayward, S. W. Kim, H. Lee, *Phys. Rev. D* **65**, 064003 (2002)
[44] S. A. Hayward, gr-qc/0203051 (2002)
[45] V. Frolov, M. Snajdr, D. Stojkovic, *Phys. Rev. D* **68**, 044002 (2003), gr-qc/0304083
[46] F. Rahaman, M. Kalam, M. Sarker, and K. Gayen, *Phys. Lett. B* **633**, 161 (2006)
[47] M. Thibeault, C. Simeone, and E. Eiroa, gr-qc/0512029 (2005)
[48] J. P. S. Lemos, F. S. N. Lobo, and S. Q. de Oliveira, *Phys. Rev. D* **68**, 064004 (2003)
[49] V. P. Frolov, *Phys. Rev. D* **74** 044006 (2006)
[50] V. P. Frolov, D. V. Fursaev, D. Stojkovic, *Class.Quant.Grav.* **21** 3483 (2004)
[51] D. Mateos, R. C. Myers, R. M. Thomson, *Phys.Rev.Lett.* 97, 091601 (2006)
[52] T. Albash, V. Filev, C. V. Johnson and A. Kundu, hep-th/0605088 (2006)
[53] E. Shuryak, *Prog. Part. Nucl. Phys.* **53**, 273 (2004)
[54] H. Chamblin and H. Reall, *Nucl. Phys. B* **562** (1999) 133;
K. I. Maeda and D. Wands, *Phys. Rev. D* **62**, 124009 (2000)
[55] C. Barceló and M. Visser, *J. High Energy Phys.* **10**, 019 (2000)
[56] R. Arnowit, S. Deser, C. Misner, *Phys. Rev.* **122**, 997 (1961)